\documentclass[aps, tightenlines, nofootinbib, 11pt]{revtex4-2}
\usepackage[english]{babel}
\usepackage{blindtext}
\usepackage[utf8x]{inputenc} 
\usepackage{amsmath,amssymb}
\usepackage{mathtools}
\usepackage{nccmath}
\usepackage{graphicx}
\usepackage{float, mathrsfs}
\usepackage[a4paper,top=2cm,bottom=2cm,left=2cm,right=2cm]{geometry}
\usepackage{xcolor}
\newcommand{\eqn}[2]{\begin{equation}\label{#1}#2 \end{equation}}
\newcommand{\p}{\varphi}

\begin{document}
\author{Anna Negro}
\email{negro@lorentz.leidenuniv.nl}
\author{Subodh P.~Patil}
\email{patil@lorentz.leidenuniv.nl}

\affiliation{Instituut-Lorentz for Theoretical Physics,\\Leiden University, 2333 CA Leiden, The Netherlands\\}
\date{\today}
	
\title{Hadamard Regularization of the Graviton Stress Tensor}

\begin{abstract}
We present the details for the covariant renormalization of the stress tensor for vacuum tensor perturbations at the level of the effective action, adopting Hadamard regularization techniques to isolate short distance divergences and gauge fixing via the Faddeev-Popov procedure. The subsequently derived renormalized stress tensor can be related to more familiar forms reliant upon an averaging prescription, such as the Isaacson or Misner-Thorne-Wheeler forms. The latter, however, are premised on a prior scale separation (beyond which the averaging is invoked) and therefore unsuited for the purposes of renormalization. This can lead to potentially unphysical conclusions when taken as a starting point for the computation of any observable that needs regularization, such as the energy density associated to a stochastic background. Any averaging prescription, if needed, should only be invoked at the end of the renormalization procedure. The latter necessarily involves the imposition of renormalization conditions via a physical measurement at some fixed scale, which we retrace for primordial gravitational waves sourced from vacuum fluctuations through direct or indirect observation.
\end{abstract}

\maketitle

\section{Introductory remarks}

Primordial gravitational waves offer an almost unimpeded view of the early universe\footnote{Subject to convolution with a non-trivial transfer function. Within the standard thermal history, the QCD crossover, changes in the number of relativistic species, and damping by free streaming neutrinos process different wavelengths in a calculable manner during radiation domination \cite{Weinberg:2003ur, Watanabe:2006qe, Kite:2021yoe} with subsequently negligible effects at later times \cite{Baym:2017xvh, Flauger:2017ged}.}. They inevitably come in the form of stochastic backgrounds, with readily inferred spectral dependences and characteristic frequencies that provide insight into the the physical mechanisms that sourced them \cite{Allen:1996vm, Christensen:2018iqi}. There is a key distinction to be drawn, however, from primordial gravitational waves sourced by dynamical processes involving energy and momentum transfer -- phase transitions, particle production, decay of topological defects, among others \cite{Caprini} -- and those corresponding to zero point fluctuations around some expanding background. Whereas the former involve propagating gravitons sourced by some physical process, the latter by definition represent vacuum polarization effects whose imprint on various observables is no less real, but with important distinctions worth qualifying. We elaborate further on these distinctions and their consequences for what can be meaningfully inferred from cosmological observations in a separate investigation \cite{NP}. 

The purpose of this article is to address the regularization and renormalization of the graviton stress tensor in a fully covariant formalism\footnote{An expanded discussion on the regularization and renormalization of the stress tensor for vacuum tensor perturbations in a foliation specific formulation mirroring the discussion here can be found in \cite{Negro:2024bbf}.}. We revisit the treatment of \cite{GW} which works directly from the effective action, extending the approach to explicitly compute the divergences and the counterterms required to subtract them on a cosmological background, obviating the need to explicitly compute finite contributions to the effective action as these are absorbed in the process of imposing renormalization conditions at some scale with certain caveats. The latter is worth stressing, as although this may seem like an old problem for which a corpus of literature exists, the majority of references focus on formal aspects of regularization, and are notably sparse on detail when it comes to the second, most consequential part of the renormalization procedure. 

In a nutshell, the renormalized stress energy tensor for gravitational waves can be obtained from the variation of the effective action with respect to the background metric. As straightforward a prescription as this is to state, determining the effective action and background metric that minimizes it is not trivial. In fact, on backgrounds that are not asymptotically flat nor corresponding to vacuum spacetimes, a number of caveats and distinctions apply to the very definition of these objects that warrant further elaboration.  

\subsection{Effective action and background field method preliminaries}
The one loop effective action can in principle be determined through a number of means. Functional techniques such as the heat kernel method (through which the DeWitt-Schwinger formalism can be implemented \cite{Schwinger:1951nm,DeWitt:1964mxt,DeWitt:1964oba, BD, Parker:2009uva}) have the advantage of being fully covariant and therefore particularly suited for computation on arbitrary backgrounds, albeit in Euclidean signature \cite{Vassilevich:2003xt}. The effective action consists of contributions can be classed according to whether they arise from the ultra-violet (UV) or the infra-red (IR) asymptotics of the proper time integral over the relevant kernel. The UV asymptotics correspond to well understood and straightforwardly calculable short distance divergences that are subtracted and renormalized in the usual way. The IR contributions -- to be distinguished from IR divergences\footnote{Unlike UV divergences, IR divergences admit multiple interpretations with implications ranging from harmless to severe. In all cases, their presence indicates that one has yet to arrive at a reliable computation of a well defined observable. If background field quantization remains valid (in that any putative IR divergences aren't indicating an unstable background) and the perturbative scheme is under control, they should cancel in all physically well defined observables. At one loop for example, IR divergences for the graviton two-point function and quantities derived from it can be shown to be an artifact of assuming a past infinite de Sitter phase, canceling on backgrounds corresponding to finite duration inflation \cite{Negro:2024bbf} (ibid. for an expanded discussion of the points raised in this footnote).} -- correspond to state and boundary condition dependence, as well as capturing physical effects like vacuum polarization and particle creation. They are generally non-local even in theories with a mass gap \cite{Barvinsky:2002uf}, which the usual DeWitt-Schwinger expansion fails to capture. Nevertheless, some of these contributions can be reliably computed with heat kernel methods via an extension of the DeWitt-Schwinger method when restricted to asymptotically flat geometries \cite{Barvinsky:1985an, Barvinsky:1987uw, Barvinsky:1990up, Barvinsky:1990uq}. Their status more generally is not known, nor in the case of boundary and state dependence, operationally knowable in completeness to a local observer\footnote{To invert a quote attributed to Alan Turing: differential equations are science, boundary conditions are religion.}. Related to this issue, is another important distinction worth stressing. 

In going from the Euclidean to Lorentzian signature for cosmological applications, one must take care to distinguish and extract quantities relevant to the Cauchy problem, namely, in-in currents and expectation values as opposed to the corresponding in-out quantities, both of which can be extracted from the Euclidean effective action through different choices of boundary conditions \cite{Barvinsky:1987uw, Barvinsky:1990up, Barvinsky:1990uq}. Following the notation of the former references, we first consider the following definitions for the effective background:\footnote{Where we also adopt DeWitt's condensed notation, and to avoid a proliferation of indices, the composite index $a$ can also be taken to denote a pair of spacetime indices $a := \{\mu,\nu\}$.}
\eqn{}{\p^a_{\rm F} = \frac{\langle {\rm out,vac} | \p^a |\rm in, vac \rangle}{\langle {\rm out,vac} | \rm in, vac \rangle},}  
which is of interest in applications when in and out states can be defined (e.g. scattering problems), and
\eqn{}{\p^a_{\rm IN} = \langle {\rm in,vac} | \p^a|{\rm in, vac} \rangle,}
which is of primary relevance to problems where only the initial state is specified. Both fields can be obtained from the effective equations of motion, which can be brought into the form:
\eqn{effeom}{\frac{\delta S_{\rm cl}}{\delta \p^a_{\rm F}} + J_a^{\rm F} = 0,~~~~\frac{\delta S_{\rm cl}}{\delta \p^a_{\rm IN}} + J_a^{\rm IN} = 0,}
where $S_{\rm cl}$ is the `classical action' and the $J_{a}$ are the so-called radiation currents which can be calculated to any order in $\hbar$ as we indicate shortly, and will correspond to renormalized energy momentum tensors. We note that Eq. \ref{effeom} has implicit contributions coming from the measure, which will be addressed in more detail in the following sections. 

Both the classical action and the respective radiation currents are functions of the corresponding background fields $\p^a_{\rm F}$ or $\p^a_{\rm IN}$, where the background field method has been implicitly adopted (see \cite{Abbott:1981ke} for a very clear and readable review). Through this method, one can efficiently compute the effective action as the sum of all 1PI vacuum graphs in the presence of a given background, with all internal lines corresponding to fluctuations around this background. Derivatives of the effective action with respect to the background field can then be used to construct all observables of interest, whether S-matrix elements or correlation functions depending on the context.  

In spite of the label, $S_{\rm cl}$ appearing in Eq. \ref{effeom} can also be thought of as incorporating an $\hbar$ expansion when expressed as 
\eqn{Scl}{S_{\rm cl} := S_0 + S_{\rm ct},}
where $S_0$ represents the tree level action for the classical background
\eqn{tree}{S_0 = S_{\rm EH} + S_{\rm M},}
and where and $S_{\rm ct}$ represents the counterterms needed to subtract the UV divergences that arise at any given loop order, with finite remainders that are to be fixed through renormalization conditions. In the above, $S_{\rm EH}$ is the Einstein-Hilbert action and $S_{\rm M}$ represents the matter content that sources the background expansion. The solution to Eq. \ref{effeom} defines a shifted tadpole condition, which is to say that the relevant background field gets corrected as\footnote{One of the advantages of the background field method is that it is possible to work perturbatively so that the background can remain unspecified until the very end, where one must eventually evaluate all physical quantities on-shell (and in the background gauge that defined the gauge fixing and ghost terms) \cite{Abbott:1981ke}.}
\eqn{bgs}{\p^a = \p^a_0 + \hbar \p^a_1 + ...}
where $\p^a_0$ minimizes $S_0$ in isolation, and where we've restored $\hbar$ to emphasize the nature of the expansion\footnote{In the context of inflationary cosmology, one show that the shifted background simplifies calculations of logarithmic corrections to inflationary correlators from the UV asymptotics of the effective action alone \cite{BDP}.}. We elaborate on the relevance of the shifted tadpole condition when fixing renormalization conditions at the end of the computation in more detail in section \ref{sec:4}.

Different diagrammatic rules apply when attempting to determine $J^{\rm F}_a$ or $J^{\rm IN}_a$. The radiation current $J^{\rm F}_a$ can be obtained via techniques relevant to the computation of transition amplitudes, and is given to one loop given by \cite{Barvinsky:1987uw}:
\eqn{SM}{J_{a}^{\rm F} = -\frac{i}{2} \frac{\delta^3 S_{\rm cl}}{\delta \p^a_{\rm F} \delta \p^b_{\rm F} \delta \p^c_{\rm F}}G^{c b}_{\rm F},~~~~~~~~ G^{c b}_{\rm F} = i \frac{\langle {\rm out,vac} |{ T} [\p^c\p^b] |\rm in, vac \rangle}{\langle {\rm out,vac} | \rm in, vac \rangle},}
where $G^{c b}_{\rm F}$ is the Feynman propagator. Similarly, the current $J^{\rm IN}_a$ is given to one loop by 
\eqn{SK}{J_{a}^{\rm IN} = -\frac{i}{2} \frac{\delta^3 S_{\rm cl}}{\delta \p^a_{\rm IN} \delta \p^b_{\rm IN} \delta \p^c_{\rm IN}}G^{c b}_{\rm IN},~~~~~~ G^{c b}_{\rm IN} = i \langle {\rm in,vac} |{ T}[\p^c\p^b] |\rm in, vac \rangle,}
where the latter can be evaluated as it appears, or with the full regalia of the 	Schwinger-Keldysh formalism. The two Green's functions differ in terms of their boundary conditions, although in the specific case of future and past asymptotic flatness, one has $|\rm out, vac \rangle = |\rm in, vac \rangle$ so that $G^{c b}_{\rm F} \equiv G^{c b}_{\rm IN}$.

Cosmological backgrounds, however, are not asymptotically flat in the past, nor in the future depending on the matter content. Nevertheless, for the purposes of regularization, only the short distance divergence structure of $G^{c b}_{\rm IN}$ is relevant, which is identical to that of $G^{c b}_{\rm F}$. The reason for this can be inferred from the fact that if the two Green's functions differ only in their boundary conditions, completeness dictates that the short distance modes of the two vacua must be related to each other by a Bogoliubov rotation that tends to zero for short wavelengths, otherwise one would represent an infinite energy excitation relative to the other (see also \cite{BD} for an expanded discussion on this point). We stress this point as it offers us the possibility to adapt computations that make use of Feynman Green's functions for the purposes of identifying the local counterterms necessitated by the subtraction procedure \cite{GW}. On the other hand, the long wavelength behavior of $G^{c b}_{\rm IN}$ will certainly differ from that of $G^{c b}_{\rm F}$. However, the difference will manifest as finite and non-local terms whose precise forms one needn't calculate beyond the scale factor dependence of the various contributions, as these will be absorbed in the process of fixing renormalization conditions, which we detail further.

A final and no less consequential clarification is necessitated by the question of whether we are obliged to work with the on, or off-shell formulation of the effective action in our computations. To one loop, the former can be obtained by expanding the action to quadratic order in fluctuations and evaluating the resulting functional determinant as a function of the background:
\eqn{sea}{\Gamma_1 = \frac{i}{2} \ln \det \left\{\frac{\delta^2 S_{\rm cl} [\p^a ]}{\delta \p_1^b  \delta \p_1^c }\right\},}
where $\p^a$ is the relevant background field, and $\p^a_1$ denote fluctuations around it. The result of differentiating the above with respect to the background yields the radiation currents Eqs. \ref{SM} and \ref{SK}, whose contractions make explicit that we are differentiating the one loop 1PI vacuum graph \cite{Abbott:1981ke}. Unlike the classical action, however, Eq. \ref{sea} is not a scalar and is dependent on how one parameterizes field space in addition to also depending on the background field gauge in the presence of gauge symmetries. 

Instead, an effective action that doesn't suffer from these drawbacks was arrived at by Vilkovisky and DeWitt by working covariantly in field space and writing down the equivalent of the functional determinant of the field covariant second variational derivative of the action:   
\eqn{vdw}{\Gamma^{\rm VdW}_1 = \frac{i}{2} \ln \det \left\{\frac{\delta^2 S_{\rm cl} [\p^a ]}{\delta \p_1^b  \delta \p_1^c } - \Gamma^d_{b c}[\p^a] \frac{\delta S_{\rm cl} [\p^a ]}{\delta \p_1^d } \right\},}
where $\Gamma^d_{b c}$ is the connection on field space. When the background field $\p^a$ minimizes $S_{\rm cl}$ (i.e. one is working on shell) the two forms are equivalent. However, the two forms will in general differ for any quantity obtained from differentiating the effective action when the field space connection is non-vanishing even when evaluated on shell. Therefore the renormalized stress tensor obtained from the Vilkovisky-DeWitt effective action will have additional contributions relative to the stress tensor obtained from the `standard theory'. However, the difference is only in terms of additional finite contributions, which moreover vanish for vacuum contributions on maximally symmetric spacetimes and for one particle states on Minkowski space \cite{GW}. On a general Friedman-Robertson-Lema\^itre-Walker (FRLW) background, with a homogeneous and isotropic fluid sourcing the background expansion, these are non-vanishing and contribute additional finite contributions that depend on the quantum state. 

Nevertheless, the procedure we follow allows us to work with the standard form of the effective action, as the divergences that need to be regularized are unaffected by the Vilkovisky-DeWitt correction term, and any finite contributions are absorbed by the process of fixing renormalization conditions. What is crucial for the subtraction process, however, is the identification of the scale factor dependence of the various finite and state dependent terms, for which we determine recursion relations with initial coefficients that can be fixed with a sufficient number of measurements at the renormalization scale. 

\subsection{Outline, scope, and synopsis}
Having informed ourselves of the necessary caveats, we can now outline the treatment to follow. We begin in section \ref{sec:2} with a discussion of the procedure by which the stress tensor associated with vacuum tensor perturbations can be derived directly from the standard effective action, addressing how the more familiar forms can be obtained from direct variation of the tree level contributions and then Brill-Hartle averaging. Sections \ref{sec:3} and \ref{sec:4} are concerned with the Hadamard regularization procedure and subsequent renormalization, in which we subtract divergences and discuss the procedure by which physical observations can be used to fix renormalization conditions, absorbing finite contributions whose explicit forms are not needed beyond their scale factor dependence in an FRLW foliation. We revisit all the important caveats and distinctions detailed above as we encounter them, and offer our summarizing thoughts in the conclusions with technical details deferred to the appendices. 

What follows adopts established techniques towards the specific problem of computing the renormalized stress tensor associated with the stochastic background of primordial gravitational waves on an FLRW background. One does not have the luxury of not following through this process to completion if one is interested in making contact with cosmological observations. As we recap in the conclusions, the literature is replete with computations for the energy density associated with the stochastic background of vacuum tensor perturbations that exhibit cutoff dependences. Moreover, these have regularized forms of the stress tensor with an intrinsic (time dependent) scale separation invoked for an averaging prescription, and therefore unsuited for the purposes of renormalization and leading to potentially unphysical conclusions when taken on face value. We refer the reader to \cite{Negro:2024bbf} for a discussion of how this process resolves in a foliation specific formulation\footnote{\label{fnIR}Where it can be shown that the a stress tensor for gravitational waves not reliant on an averaging prescription and valid for all wavelengths can be regularized, the necessary counterterms identified -- the step at which previous treatments would have identified their computations going off course -- and IR divergences canceled on backgrounds that transitioned into radiation domination from a period of finite duration (as opposed to past infinite) inflation.}, where it is stressed that any attempts to infer bounds on the number of relativistic species in the early universe from primordial gravitational waves is inextricable from the process of renormalization. What follows is an reexamination of this procedure in a fully covariant formalism, with a more detailed exploration of its implication for cosmological and astrophysical implications deferred to a separate investigation \cite{NP}.

\section{Vacuum stress energy from the effective action}\label{sec:2}
Our treatment proceeds from the standard effective action via the background field method. We expand the classical Einstein-Hilbert action to quadratic order in perturbations $h_{\mu \nu}$ around some background, defined as 
\begin{equation}
	\bar{g}_{\mu \nu}=g_{\mu \nu}+h_{\mu \nu},
\end{equation}
where $g_{\mu \nu}$ is the background metric which we leave unspecified for the time being. To second order, one can show that \cite{GW, Barth:1983hb}
\begin{eqnarray}\label{eq:s2}
	\nonumber	S^{(2)}&= \frac{\kappa^{2}}{2} \int d^4 x \sqrt{-g} & \left[ \frac{1}{2}h_{\rho \sigma}\Box h^{\rho \sigma}  + h \nabla^{\rho} \nabla^{\sigma}h_{\sigma \rho } + \nabla^{\alpha}h_{\alpha}{}^{ \rho }\nabla^{\sigma}h_{\sigma \rho } -\frac{1}{2} h \Box h + R^{\beta}{}_{\rho \alpha \sigma} h_{ \beta}{}^{\alpha} h^{\rho \sigma} + h_{\alpha}{}^{ \rho}h^{\alpha \sigma}R_{ \rho \sigma}\right.\\
	&& \left.  -h  h^{\rho \sigma}R_{ \rho \sigma} - \frac{1}{2} h^{\rho \sigma}h_{\rho \sigma} R + \frac{1}{4} h h R  \right],
\end{eqnarray}
where we've defined $\kappa^2 \equiv \frac{1}{8 \pi G_{N}}$. The stress tensor for gravitational waves is obtained by variation with respect to the background metric -- a process that is equivalent to perturbing the Einstein equations to second order and bringing the quadratic terms over to the other side to act as a source for the background. 

If one is only interested in backgrounds of gravitational waves with a spectrum of bounded support and comprised wavelengths and periods much shorter than the background curvature scale\footnote{\label{fn1}That is, when $\lambda, \omega^{-1} \ll 2\pi \mathscr R$, where $\mathscr R^{-2}$ is the typical magnitude of the non-vanishing components of the background Riemann tensor.} -- as with most astrophysical applications -- a number of approximations and simplifications are possible. Firstly, one can neglect terms containing derivatives of the background relative to derivatives of the tensor perturbations, and as a result, discard terms proportional to the curvature scalar and Ricci tensor and freely commute covariant derivatives. Moreover, one can invoke a spatial averaging over all physical wavenumbers greater than the inverse radius of background curvature (itself, a time dependent criterion on cosmological backgrounds) and integrate by parts within the averaging integral \cite{Brill:1964zz, Maggiore}.  Doing so results in the simplified expression \cite{Misner:1973prb}
\eqn{MTW}{T_{\mu \nu}^{\mathrm{gw}}=\frac{1}{32 \pi G_N}\left\langle \bar h_{\alpha \beta; \mu} \bar h_{~~; \nu}^{\alpha \beta}-\frac{1}{2} \bar h_{; \mu} \bar h_{; \nu}- \bar h^{\alpha \beta}{ }_{; \beta} \bar h_{\alpha \mu; \nu}- \bar h_{~~; \beta}^{\alpha \beta}\bar h_{\alpha \nu; \mu}\right\rangle_{\rm BH}}
where subscript on the angled brackets denotes spatial averaging, semi-colons denote covariant derivatives with respect to the background metric, and where we've defined
\eqn{Isc}{\bar  h_{\mu\nu} = h_{\mu\nu} - \frac{1}{2}g_{\mu\nu}h.}
Eq. \ref{MTW} is covariantly conserved and gauge invariant up to correction terms that are negligible for all wavelengths shorter than the averaging scale. Further specifying transverse-traceless gauge results in the Isaacson form of the stress tensor \cite{Isaacson} (see also \cite{Maccallum:1973gf,Stein:2010pn}):
\begin{equation}\label{BH averaged Tmunu}
	T_{\mu \nu}^{\rm gw, Isc} = \frac{1}{32 \pi G_{N}}\langle  h_{\rho \sigma; \mu} h^{\rho \sigma}_{~~;\nu} \rangle_{\rm BH}.
\end{equation}
The expression above is widely taken as the starting point for determining the energy density associated with stochastic backgrounds of primordial origin (see e.g. \cite{Caprini}). While this is certainly valid for spectra of bounded support sourced by some physical production mechanism (and hence sub-horizon), taking this formula on face value for modes with wavelengths greater than the domain of validity defined by the averaging scale ought to be treated with caution. This caution should be amplified when one encounters divergences that need to be regularized, as is the case for stochastic backgrounds associated with vacuum tensor perturbations given the prior scale separation inherent in the definition of Eqs. \ref{MTW} and \ref{Isc}. Reverting back to the approximately gauge invariant form Eq. \ref{MTW} won't help matters either, as any computation at any loop order implicit integrates over all scales and will eventually run afoul of this approximation. Appeals to ad hoc IR cutoffs should be viewed through the lens that if one recovers divergences and gauge dependence in one's answer as this cutoff is removed, one has not arrived at a physically reliable answer. So does one proceed instead?

At this point, an important aside is due: although one might find statements in the literature that questions the utility of even defining gravitational waves with wavelengths grater than the background curvature scale\footnote{In section 35.7 of \cite{Misner:1973prb} for example, one finds the statement that ``One must always have $\mathscr{A} \ll 1$ as well as $\lambda \ll 2\pi \mathscr R$ if the concept of gravitational wave is to make any sense", where $\mathscr A$ is the dimensionless amplitude of the gravitational wave and $\mathscr R^{-2}$ is as defined in footnote \ref{fn1} (see also \cite{Maggiore:2007ulw, Maccallum:1973gf, Stein:2010pn, Caprini} for more detailed discussions of this point).}, this would nominally be at odds with the premise of many computations. It is also add odds with observations: gravitational waves from mergers of binary black hole systems have been observed \cite{KAGRA:2021vkt, LIGOScientific:2020iuh, LIGOScientific:2020ufj}. Moreover, the search for primordial gravitational waves is premised on the fact that they induce local quadrupolar anisotropies in the density field of the primordial plasma at all scales, resulting in a signature B-mode polarization pattern \cite{Kamionkowski:1996ks, Seljak:1996gy}. Both situations feature gravitational waves with wavelengths comparable to or greater than the background curvature radius at some point -- the peak frequency emitted from a merger corresponding to wavelengths commensurate to the Schwarzschild radius, and primordial tensor fluctuations having crossed the Hubble radius before sourcing local anisotropies. 
 
Clearly, nature tells us that the notion of tensor perturbations with wavelengths longer than than the background curvature radius has to make sense. By general covariance and the Bianchi identities that follow as a corollary, one must also be able to identify a conserved rank two tensor that plays the role of a stress tensor from direct perturbation of the equations of motion. That one can do so with minimal fuss by simply undoing the averaging prescription and avoiding neglecting any terms arising from the direct variation of Eq. \ref{eq:s2} can be found in \cite{Negro:2024bbf}\footnote{Although the notion of long wavelength tensor perturbations certainly makes sense, assigning an energy density to sufficiently long wavelength perturbations does come with caveats, although essentially semantic in nature and rendered moot in the extraction of quantities that imprint on observables -- see Section IV of \cite{Negro:2024bbf} for an expanded discussion.}. It is also the premise of the computation in \cite{GW} that is retraced in what follows, which stresses that Eq. \ref{eq:s2} is to be viewed as the action for a massless spin two degree of freedom on a curved background, whose regularization and renormalization proceeds via established prescriptions. 

We proceed from the unadulterated definition for the stress tensor for gravitational waves:
\begin{equation}\label{eq:Tmunu gw 1}
	T_{\mu \nu}^{\rm gw} = -  \frac{2}{\sqrt{-g}}  \left\langle \frac{\delta \left(- S_{\rm gw} \right)}{\delta g^{\mu \nu}} \right\rangle_{\rm in, in}
\end{equation}
where the in-in expectation value implicitly traces over some initial density matrix. When this state is taken as the adiabatic vacuum, one obtains the stress tensor for vacuum tensor perturbations. 

In order to define $S_{\rm gw}$ we must first consider the process of gauge fixing, for which de Donder gauge presents a particularly efficient choice. We proceed via the Faddeev-Popov method \cite{Faddeev:1967fc} and add a gauge breaking term which fixes the chosen gauge condition to Eq. \ref{eq:s2}, defined as $ \nabla_{\mu} h^{\mu \nu} = \frac{1}{2}\nabla_\nu h$, along with a ghost term that accounts for the measure factor induced by gauge fixing:
\begin{eqnarray}\label{eq:sgb sgh}
		&S_{\rm gb}= &- \frac{\kappa^{2}}{2} \int d^4 x \sqrt{-g}  \left( \nabla^{\mu} h_{\mu \nu}-\frac{1}{2} \nabla_{\nu} h\right)\left(\nabla^{\alpha} h_{\alpha}{}^{\nu}-\frac{1}{2} \nabla^{\nu} h \right),\\
		&S_{\rm gh}= & \frac{\kappa^{2}}{2} \int d^4 x \sqrt{-g}  \left[   \bar{\eta}^{\mu}\left(g_{\mu \nu} \square-R_{\mu \nu}\right) \eta^{\nu} \right].
\end{eqnarray}
In the above, $\eta^\rho$ represents the ghost field that accounts for the residual gauge freedom by subtracting the spurious degrees of freedom from the action $S^{(2)}$. Consequently, the gauge fixed action for the gravitational sector is given by:
\begin{eqnarray}\label{eq:sgw}
		S_{\rm gw}&=& S^{(2)} + S_{\rm gb} + S_{\rm gh}\\
		\nonumber& =&\frac{\kappa^{2}}{2} \int d^4 x \sqrt{-g}  \left[ \frac{1}{2}h_{\rho \sigma}\Box h^{\rho \sigma} -\frac{1}{4} h \Box h + R^{\beta}{}_{\rho \alpha \sigma} h_{ \beta}{}^{\alpha} h^{\rho \sigma} + h_{\alpha}{}^{ \rho}h^{\alpha \sigma}R_{ \rho \sigma} -h  h^{\rho \sigma}R_{ \rho \sigma}   - \frac{1}{2} h^{\rho \sigma}h_{\rho \sigma} R  \right.\\
		\nonumber && \left.   + \frac{1}{4} h h R  +  \bar{\eta}^{\mu}\left(g_{\mu \nu} \square-R_{\mu \nu}\right) \eta^{\nu}\right].
\end{eqnarray}
In this way, starting with the action Eq. \ref{eq:s2} for a rank-2 symmetric tensor field nominally consisting of ten degrees of freedom, we obtain the action of a massless spin-2 particle $S_{\rm gw}$ with only two propagating degrees of freedom.
From Eq. \ref{eq:Tmunu gw 1}, the stress energy tensor of vacuum tensor perturbations is given by the sum of contributions
\begin{equation}\label{eq:Tmunu gw 2}
	T_{\mu \nu}^{\rm gw}  = - \frac{2}{\sqrt{-g}}  \left\langle \frac{\delta }{\delta g^{\mu \nu}}\left( - S_{\rm gr} - S_{\rm gh} \right) \right\rangle_{\rm in, in} = T_{\mu \nu}^{\rm gr}+T_{\mu \nu}^{\rm gh},
\end{equation}
where we've defined $S_{\rm gr}\equiv S^{(2)} + S_{\rm gb}$. The angled brackets above denotes the time ordered in-in correlation function $\langle ... \rangle := \langle {\rm in,vac} |{\rm } T [...] | {\rm in,vac} \rangle$, which inevitably exhibits divergences for field bilinears in the coincident limit, the regularization of which we turn to next.

\section{Regularization} \label{sec:3}
The regularization of the stress tensor for any propagating degree of freedom on a general background must proceed with care, all the more so when gauge redundancies are present. Although we proceed to do so in a covariant manner, one can also contemplate working directly at the level of the stress tensor obtained by variation of Eq. \ref{eq:s2}, gauge fixing by hand, and then imposing the scalar vector tensor decomposition to extract the stress tensor for the propagating spin two polarizations when evaluated as an expectation value \cite{Negro:2024bbf}. This procedure can be related to the covariant method detailed below by a series of Ward identities that we return to further on. In what follows, we proceed to regularize the divergences encountered in the evaluation of Eq. \ref{eq:Tmunu gw 2} by adopting Hadamard regularization techniques \cite{GW, Hadamard 1, Hadamard 2}, which are an extension of the covariant point-splitting method\footnote{cf. \cite{Schwinger:1951nm, Newton:1993mb} for applications of point splitting to gauge theories on flat space, and \cite{II(2),Brown:1986tj, Balakumar:2019djw, Decanini:2005gt, Belokogne:2015etf} for applications of Hadamard techniques to scalar, vector and fermionic degrees of freedom on curved backgrounds.}.

\subsection{Hadamard point splitting} \label{sec:3.1}

The point-split version of a tensor $U^{\mu\nu} (x)$ is defined as the coincidence limit of the bitensor $U^{\mu\nu^{\prime}} (x, x^{\prime})$ defined in a neighbourhood of $x^{\mu}$ 
\begin{equation}\label{def point splitting}
	U^{\mu\nu} (x)= \lim_{\sigma^\mu  \to 0} U^{\mu\nu^{\prime}} (x, x^{\prime}),
\end{equation}
where primed indeces refer to the point $x^{\mu^{\prime}}$ and $\sigma^\mu $ is the geodesic distance between $x^{\mu}$, and $x^{\mu^{\prime}}$. Doing so allows us to isolate the divergent from finite contributions to Eq. \ref{eq:Tmunu gw 2} with $\sigma^{\mu} $ as the UV regulator. Hadamard regularization of the effective action proceeds through the intermediary of the Feynman propagators (in the notation of \cite{GW}):
\begin{eqnarray}\label{eq:Hadamard F propagators}
		G^{\mu \nu \alpha^{\prime} \beta^{\prime}}\left(x, x^{\prime}\right)&=&\frac{i}{32 \pi G_{ N}} \frac{\langle\psi| T\left(h^{\mu \nu}(x) h^{\alpha^{\prime} \beta^{\prime}}\left(x^{\prime}\right)\right)|\psi\rangle}{\langle\psi \mid \psi\rangle}\\
		\nonumber &=&\frac{i}{8 \pi^{2}}\left[\frac{\Delta^{1 / 2}}{\sigma+i \varepsilon}\left(g^{\alpha^{\prime} (\mu}g^{\nu) \beta^{\prime}} \right)+V^{\mu \nu \alpha^{\prime} \beta^{\prime}} \ln (\Lambda^2(\sigma+i \varepsilon))+W^{\mu \nu \alpha^{\prime} \beta^{\prime}}\right]\\
		\nonumber \tilde{G}^{\mu \alpha^{\prime}}\left(x, x^{\prime}\right)&=&\frac{i}{32 \pi G_{ N}} \frac{\langle\psi| T\left(\bar{\eta}^{\mu}(x) \eta^{\alpha^{\prime}}\left(x^{\prime}\right)\right)|\psi\rangle}{\langle\psi \mid \psi\rangle}\\
		\nonumber &=&\frac{i}{8 \pi^{2}}\left[\frac{\Delta^{1 / 2}}{\sigma+i \varepsilon} g^{\mu \alpha^{\prime}}+\tilde{V}^{\mu \alpha^{\prime}} \ln (\Lambda^2(\sigma+i \varepsilon))+\tilde{W}^{\mu \alpha^{\prime}}\right],
\end{eqnarray}
where $\Lambda$ is some arbitrary mass scale so that the argument of the logarithms are dimensionless, and primed indices are geodesically transported from $x^{\mu^\prime}$ to $x^{\mu}$ by using the bivector of parallel displacement $g^{\alpha^{\prime}}{}_{\alpha}$, defined by the differential equation \cite{DeWitt:1964mxt, DeWitt:1960fc}
\begin{equation}
	\nabla_\rho g_{\alpha^{\prime} \beta} \nabla^\rho \sigma = 0,
\end{equation}
with the boundary condition 
\begin{equation}
	\lim_{x \to x'} g_{\alpha^{\prime} \beta} (x, x') = g_{\alpha \beta} (x).
\end{equation}
The $i\varepsilon$ appearing above is characteristic of the Feynman propagator. Within the present context it is more convenient to work with the Hadamard Green's functions, for which we follow \cite{Belokogne:2015etf} in deriving their representations from the Hadamard form of the Feynman propagator. By using the identities
\eqn{}{\frac{1}{\sigma +i\varepsilon} = \mathcal{P} \frac{1}{\sigma} - i \pi \delta(\sigma), \quad \ln (\sigma +i\varepsilon) = \ln |\sigma | + i \pi \Theta(- \sigma), }
where $\mathcal{P}$ and $\Theta$ denote the Cauchy principal value and the Heaviside theta function respectively, we can rewrite Eq. \ref{eq:Hadamard F propagators} as:
\eqn{}{G_F^{ab}(x,x^\prime) = G_A^{ab}(x,x^\prime) + \frac{i}{2} G^{ab} (x,x^\prime), }
where 
\begin{eqnarray}\label{eq:averaged  propagators}
	G_A^{\mu \nu \alpha^{\prime} \beta^{\prime}}\left(x, x^{\prime}\right) &=&\frac{1}{8 \pi}\left[\Delta^{1 / 2} \left(g^{\alpha^{\prime} (\mu}g^{\nu) \beta^{\prime}} \right) \delta(\sigma) -V^{\mu \nu \alpha^{\prime} \beta^{\prime}}  \Theta(- \sigma) \right]\\
	\nonumber \tilde{G}_A^{\mu \alpha^{\prime}}\left(x, x^{\prime}\right)&=&\frac{1}{8 \pi}\left[\Delta^{1 / 2}  g^{\mu \alpha^{\prime}} \delta(\sigma)-\tilde{V}^{\mu \alpha^{\prime}}  \Theta(- \sigma) \right],
\end{eqnarray}
is the average of the andvanced and retarded Green's functions, and 
\begin{eqnarray}\label{eq:Hadamard  propagators}
	G^{\mu \nu \alpha^{\prime} \beta^{\prime}}\left(x, x^{\prime}\right) &=&\frac{1}{4 \pi^{2}}\left[\frac{\Delta^{1 / 2}}{\sigma}\left(g^{\alpha^{\prime} (\mu}g^{\nu) \beta^{\prime}} \right)+V^{\mu \nu \alpha^{\prime} \beta^{\prime}} \ln (\Lambda^2 \sigma)+W^{\mu \nu \alpha^{\prime} \beta^{\prime}}\right]\\
	\nonumber \tilde{G}^{\mu \alpha^{\prime}}\left(x, x^{\prime}\right)&=&\frac{1}{4 \pi^{2}}\left[\frac{\Delta^{1 / 2}}{\sigma} g^{\mu \alpha^{\prime}}+\tilde{V}^{\mu \alpha^{\prime}} \ln (\Lambda^2\sigma)+\tilde{W}^{\mu \alpha^{\prime}}\right],
\end{eqnarray}
is the Hadamard Green's function.

The state $| \psi\rangle$ appearing in Eq. \ref{eq:Hadamard propagators} is somewhat circularly defined as any quantum state -- the Hadamard state -- such that the short distance divergence structure is of the forms indicated in the square brackets, where $\sigma= \frac{1}{2} \sigma_\mu \sigma^\mu $ denotes the square of the geodesic distance between $x^{\mu}$ and $x^{\mu^{\prime}}$, $\Delta$ is the Van Vleck-Morette determinant and the bitensors $V^{\mu \nu \alpha^{\prime} \beta^{\prime}}$, $W^{\mu \nu \alpha^{\prime} \beta^{\prime}}$, $\tilde{V}^{\mu \alpha^{\prime}} $ and $\tilde{W}^{\mu \alpha^{\prime}}$ are smooth functions in the limit $\sigma \to 0$ of the form:
\begin{eqnarray}\label{eq:UW expansions}
		&V^{\mu \nu \alpha^{\prime} \beta^{\prime}}=\sum_{n=0}^{\infty} V^{\mu \nu \alpha^{\prime} \beta^{\prime}}_{n} \sigma^{n} \hspace{1.5cm}
		&W^{\mu \nu \alpha^{\prime} \beta^{\prime}}=\sum_{n=0}^{\infty} W^{\mu \nu \alpha^{\prime} \beta^{\prime}}_{n} \sigma^{n}\\
		\nonumber &\tilde{V}^{\mu \alpha^{\prime}}=\sum_{n=0}^{\infty} \tilde{V}^{\mu \alpha^{\prime}}_{n} \sigma^{n} \hspace{1cm}
		&\tilde{W}^{\mu \alpha^{\prime}}=\sum_{n=0}^{\infty} \tilde{W}^{\mu \alpha^{\prime}}_{n} \sigma^{n}.
\end{eqnarray}
It is to be stressed that the bitensors $V_n^{\mu \nu \alpha^{\prime} \beta^{\prime}}$ and $\tilde{V}_n^{\mu \alpha^{\prime}}$ depend only on the local geometry, whereas the bitensors  $W_n^{\mu \nu \alpha^{\prime} \beta^{\prime}}$ and $\tilde{W}_n^{\mu \alpha^{\prime}}$ depend on the boundary conditions and the precise choice of the state $|\psi\rangle$. Therefore, any difference between evaluating Eq. \ref{eq:Tmunu gw 2} with in-in expectation values as opposed to their in-out counterparts as encoded in the Feynman propagators of Eq. \ref{eq:Hadamard propagators} will manifest only in differences in the finite contributions $W_n^{\mu \nu \alpha^{\prime} \beta^{\prime}}$ and $\tilde{W}_n^{\mu \alpha^{\prime}}$. It is precisely these finite contributions that are moot to any local observer, being absorbed by renormalized couplings in the effective action through the process of imposing renormalization conditions.   

Each of the $V_n^{\mu \nu \alpha^{\prime} \beta^{\prime}}$, $W_n^{\mu \nu \alpha^{\prime} \beta^{\prime}}$, $\tilde{V}_n^{\mu \alpha^{\prime}} $ and $\tilde{W}_n^{\mu \alpha^{\prime}}$ bitensors can be rewritten in the form of a covariant Taylor expansion for $x^\mu$ in the neighbourhood of $x^{\mu^{\prime}}$:
\begin{eqnarray}\label{eq:Taylor expansions}
		&& V^{\mu \nu \alpha^{\prime} \beta^{\prime}}_{n} = g^{\alpha^{\prime} }{}_{\alpha} g^{\beta^{\prime} }{}_{\beta} \left[v_n^{\mu \nu \alpha \beta}+v_{n}^{\mu \nu \alpha \beta}{}_{\gamma} \sigma^{\gamma} +\frac{1}{2}  v_{n }^{\mu \nu \alpha \beta}{}_{\gamma \tau} \sigma^{\gamma}  \sigma^{\tau} +... \right], \\
		\nonumber && \tilde{V}^{\mu  \alpha^{\prime}}_{n} = g^{\alpha^{\prime} }{}_{\alpha}  \left[\tilde{v}_n^{\mu  \alpha }+\tilde{v}_{n }^{\mu  \alpha }{}_{\gamma} \sigma^{\gamma} +\frac{1}{2}  \tilde{v}_{n }^{\mu  \alpha }{}_{\gamma \tau} \sigma^{\gamma}  \sigma^{\tau} +... \right],
\end{eqnarray}
and similarly for $W_n^{\mu \nu \alpha^{\prime} \beta^{\prime}}$ and $\tilde{W}_n^{\mu \alpha^{\prime}}$.

The Hadamard Green's functions as expressed in Eq. \ref{eq:Hadamard propagators} facilitate the regularization of Eq. \ref{eq:sgw} in that they can be viewed as the point split expression: 
\begin{eqnarray}\label{eq:Actions 2.1}
		&\left\langle S_{\rm gr} \right\rangle   &=\lim _{\sigma^\mu  \to 0}\left\{\int d ^ { 4 } x \sqrt { - g } \left[\left(-\frac{1}{2} g_{\rho \alpha^{\prime}} g_{\sigma \beta^{\prime}}+\frac{1}{4} g_{\rho \sigma} g_{\alpha^{\prime} \beta^{\prime}}\right) \nabla_{\tau} \nabla^{\tau^{\prime}}G^{\rho \sigma \alpha^{\prime} \beta^{\prime}}\right.\right.\\
		&&\nonumber \left.\left.+ \left( R_{\alpha^{\prime} \rho \beta^{\prime} \sigma}+g_{\beta^{\prime} \sigma} R_{\rho \alpha^{\prime}}-g_{\alpha^{\prime} \beta^{\prime}} R_{\rho \sigma}-\frac{1}{2} R \: g_{\rho \alpha^{\prime}} g_{\sigma \beta^{\prime}} +\frac{1}{4} R \:g_{\rho \sigma} g_{\alpha^{\prime} \beta^{\prime}} \right)  G^{\rho \sigma \alpha^{\prime} \beta^{\prime}} \right] \right\}\\ \label{eq:Actions 2.2}
		&\left\langle S_{\rm gh}  \right\rangle&=\lim _{\sigma^\mu  \to 0}\left\{ \int d^{4} x \sqrt{-g}\left[-g_{\mu \alpha^{\prime}}\nabla_{\tau} \nabla^{\tau^{\prime}}\tilde{G}^{\mu \alpha^{\prime}}-R_{\mu \alpha^{\prime}}\tilde{G}^{\mu \alpha^{\prime}}\right] \right\}.
\end{eqnarray}
Recalling that the divergence structure of the Hadamard Green's functions is completely captured by the terms containing the Van Vleck-Morette determinant $\Delta$, and the bitensors $V^{\mu \nu \alpha^{\prime} \beta^{\prime}}$ and $\tilde{V}^{\mu \alpha^{\prime}} $, the divergent part of the gravitational sector of the effective action Eq. \ref{eq:sgw} must be of the form
\begin{equation}\label{eq:regularized action 1}
	\left\langle S_{\rm gw}  \right\rangle_{\rm div}  \sim \lim_{\sigma^\mu \to 0} \int d^{4} x \sqrt { - g } \left[ \gamma_1(\sigma) R + \gamma_2 (\sigma) R^2 + \gamma_3 (\sigma) R^{\mu \nu} R_{\mu \nu} + \gamma_4(\sigma)\square R\right].
\end{equation}
We compute the coefficients $\gamma_1(\sigma)$, $\gamma_2(\sigma) $, $\gamma_3(\sigma)$, and $\gamma_4(\sigma)$ in the next subsection, from which one can immediately identify the counterterms required to subtract them. 

\subsection{Counterterms} \label{sec:3.2}
The process of determining the counterterms needed to regularize the effective action begins with rewriting the coincidence limits as: 
\begin{equation}\label{eq:endpoint ghost expansion}
	\begin{aligned}
		&\lim_{\sigma^{\mu} \to 0} R_{\mu \alpha^{\prime}}\tilde{G}^{\mu \alpha^{\prime}}=R_{\mu \alpha}  \lim_{\sigma^{\mu} \to 0} g_{\alpha^{\prime}}{}^{\alpha} \tilde{G}^{\mu \alpha^{\prime}}
	\end{aligned}
\end{equation}
so that that the tensors contracted with the Hadamard Green's functions are tensors in $x^{\mu}$ and scalars in $x^{\mu^\prime}$. Furthermore, as the coincidence limit depends on the path by which $\sigma^{\mu}$ approaches zero, it is necessary to specify a path-averaging procedure. Following Ref. \cite{Adler:1976jx}, we use the so-called \textit{elementary averaging procedure}, whereby one makes the replacements:
\begin{eqnarray}\label{eq:average}
		 \sigma_{\lambda} \sigma_{\mu} &\rightarrow & \frac{1}{4} g_{\lambda \mu} \sigma_{\rho} \sigma^{\rho}={ }_{2}^{1} g_{\lambda \mu} \sigma, \\
	\nonumber	 \sigma_{\lambda} \sigma_{\mu} \sigma_{\gamma} \sigma_{\delta} & \rightarrow & \frac{1}{6} \sigma^{2}\left(g_{\gamma \delta} g_{\lambda \mu}+g_{\gamma \lambda} g_{\delta \mu}+g_{\gamma \mu} g_{\delta \lambda}\right), \\
	\nonumber	 \sigma_{\alpha} \sigma_{\beta} \sigma_{\lambda} \sigma_{\mu} \sigma_{\gamma} \sigma_{\delta} & \rightarrow &(1 / 24) \sigma^{3}\left[g_{\alpha \beta}\left(g_{\lambda \mu} g_{\nu \delta}+g_{\lambda \gamma} g_{\mu \delta}+g_{\lambda \delta} g_{\mu \nu}\right)+g_{\alpha \lambda}\left(g_{\beta \mu} g_{\gamma \delta}+g_{\beta \gamma} g_{\mu \delta}+g_{\beta \delta} g_{\mu \gamma}\right)\right.\\
		\nonumber &&+g_{\alpha \mu}\left(g_{\beta \lambda} g_{\gamma \delta}+g_{\beta \nu} g_{\lambda \delta}+g_{\beta \delta} g_{\lambda \gamma}\right) +g_{\alpha \gamma}\left(g_{\beta \lambda} g_{\mu \delta}+g_{\beta \mu} g_{\lambda \delta}+g_{\beta \delta} g_{\lambda \mu}\right) \\
	\nonumber	&&\left.+g_{\alpha \delta}\left(g_{\beta \lambda} g_{\mu \nu}+g_{\beta \mu} g_{\lambda \gamma}+g_{\beta \gamma} g_{\lambda \mu}\right)\right].
\end{eqnarray}
The singularity structure is then extracted by expanding at the endpoints and iteratively solving the equation of motion of the propagator to find the Taylor coefficients of $V^{\mu \nu \alpha^{\prime} \beta^{\prime}}$ and $\tilde{V}^{\mu \alpha^{\prime}} $ (see appendix \ref{Appendix A} for more details). From this, the divergent contributions to $S_{\rm gh}$ are found to be
\begin{equation}\label{regularized ghost action}
		\left\langle S_{\rm gh}\right\rangle_{\rm div}  =  \frac{1}{4 \pi^2} \lim _{\sigma^{\mu} \rightarrow 0}  \int d^{4} x \sqrt { - g } \left[ - \frac{3}{\sigma} R+\ln (\Lambda^2\sigma) \left( R_{\mu \nu} R^{\mu \nu}+\frac{1}{6}R^{2}-\frac{1}{24} R_{\mu \nu \rho  \sigma} R^{\mu \nu \rho \sigma}+ \frac{5}{12} \Box R \right)  \right], 
\end{equation}
whereas the divergent contributions to $S_{gw}$ are given by:
\begin{equation}\label{regularized graviton action}
		\left\langle S_{\rm gr}\right\rangle_{\rm div}  = \frac{1}{4 \pi^2} \lim _{\sigma^{\mu} \rightarrow 0} \int d^{4} x \sqrt { - g } \left[-\frac{7}{6\sigma}  R + \ln (\Lambda^2\sigma) \left( R_{\mu \nu} R^{\mu \nu} -\frac{1}{4} R^2 -\frac{13}{24} \Box R\right) \right].
\end{equation}
By use of the Bianchi identity and the Gauss-Bonnet theorem, the divergent contribution of the action of vacuum tensor fluctuations results from the difference of Eq. \ref{regularized graviton action} and Eq. \ref{regularized ghost action} (accounting for the statistics of the ghost contributions), and is given by:
\begin{equation}\label{divc}
		\left\langle S_{\rm gw}\right\rangle_{\rm div}  =\frac{1}{4 \pi^2} \lim _{\sigma^{\mu} \rightarrow 0}  \int d^{4} x \sqrt { - g } \left[\frac{11}{6 \sigma}  R+ \ln (\Lambda^2\sigma) \left(\frac{1}{6} R_{\mu \nu} R^{\mu \nu} -\frac{11}{24} R^2 -\frac{23}{24} \Box R\right) \right],
\end{equation}
which is of the form of Eq. \ref{eq:regularized action 1}, with the $\gamma_i(\sigma)$ coefficients straightforwardly read off from the above.

\section{Renormalization} \label{sec:4}

Having identified the divergent contributions to gravitational sector of the effective action, we can now proceed to subtract them with the appropriate counterterms and absorb finite contributions through the imposition of renormalization conditions. We begin by considering all contributions to the action (cf. Eqs. \ref{Scl}, \ref{tree}, and \ref{eq:sgw}):
\begin{equation} 
\left\langle S \right\rangle = S_{\rm EH} + S_{\rm RD} + S_{\rm ct} +  \left\langle S_{\rm gw}\right\rangle, 
\end{equation} 
where the matter content that sources the background expansion $S_{\rm M}$ is now specified by $S_{\rm RD}$ to denote radiation domination, although the treatment that follows generalizes to any background. For concreteness, we work with the action formulation for a barotropic fluid expressed in terms of a derivatively coupled scalar \cite{Boubekeur:2008kn, Armendariz-Picon:2000ulo}, so that
\eqn{px}{S_{\rm RD} = \int d^4x \sqrt{-g} P^{\rm bg}(X),}  
where $P^{\rm bg}(X) = X^2$, with $X := -\frac{1}{2}g^{\mu\nu}\partial_\mu\psi\partial_\nu\psi$ reproduces the stress tensor for the background radiation fluid\footnote{A feature that remains true to all orders in perturbation theory \cite{Boubekeur:2008kn}.}, and where we've defined 
\begin{equation}\label{Sct}
		S_{\rm ct}  =\frac{1}{4 \pi^2}  \lim _{\sigma^{\mu} \rightarrow 0}\int d^{4} x \sqrt { - g } \Bigl[\alpha_1(\sigma, \mu)  R +  \alpha_2(\sigma, \mu)R_{\mu \nu}R^{\mu \nu} + \alpha_3(\sigma, \mu) R^2 + \alpha_4(\sigma, \mu) \Box R\Bigr],
\end{equation}
where $\mu$ is an arbitrary mass scale whose meaning will become clear shortly. The divergent contributions in Eq. \ref{divc} can be subtracted with the following choices for the $\alpha_i$:
\begin{eqnarray}\label{counterterms}
	\alpha_1 (\mu,\sigma)&=& -\frac{11}{6}\frac{1}{\sigma} + \alpha_1^{\rm F}(\mu), \\
	\nonumber \alpha_2(\mu, \sigma)&=& -\frac{1}{6} \log(\mu^2 \sigma) + \alpha_2^{\rm F}(\mu),  \\
	\nonumber \alpha_3(\mu, \sigma )&=& \frac{11}{24} \log(\mu^2 \sigma) + \alpha_3^{\rm F}(\mu),\\
	\nonumber \alpha_4(\mu, \sigma )&=& \frac{23}{24} \log(\mu^2 \sigma) + \alpha_4^{\rm F}(\mu),
\end{eqnarray}
where the $\alpha_i^{\rm F}$ are finite contributions that we leave unspecified for now. With this, the regularized, but yet to be renormalized action can be expressed as:
\begin{equation}\label{full action 2}
\left\langle S \right\rangle=  \int d^{4} x \sqrt{-g}\left[\frac{1}{16 \pi G(\mu)} R +\bar\alpha_2 (\mu) R_{\mu \nu} R^{\mu \nu}+ \bar\alpha_3(\mu) R^{2}+\bar\alpha_{4}(\mu) \Box R + P^{\rm bg}(X_B)\right] + \left\langle S_{\rm gw}\right\rangle_{\rm  fin},
\end{equation}
where $\left\langle S_{\rm gw}\right\rangle_{\rm fin}: = \left\langle S_{\rm gw}\right\rangle - \left\langle S_{\rm gw}\right\rangle_{\rm div}$ (cf. appendix \ref{Appendix B}), and where
\begin{eqnarray}\label{counterterms2}
	\frac{1}{16\pi G(\mu)}&=& \frac{1}{16\pi G_B} + \frac{\alpha_1^{\rm F}(\mu)}{4\pi^2}, \\
	\nonumber \bar\alpha_2(\mu) &=& \frac{1}{4\pi^2}\left[\frac{1}{6} \log\frac{\Lambda^2}{\mu^2} + \alpha_2^{\rm F}(\mu)\right],  \\
	\nonumber \bar\alpha_3(\mu) &=& \frac{1}{4\pi^2}\left[-\frac{11}{24} \log\frac{\Lambda^2}{\mu^2} + \alpha_3^{\rm F}(\mu)\right],\\
	\nonumber \bar\alpha_4(\mu) &=& \frac{1}{4\pi^2}\left[-\frac{23}{24} \log\frac{\Lambda^2}{\mu^2} + \alpha_4^{\rm F}(\mu)\right].
\end{eqnarray}
The subscripts $B$ appearing above are to denote bare quantities. 

In order to proceed, we are obliged to make use of leading order field equations to eliminate redundant higher order correction terms containing second derivatives and time derivatives of what were auxiliary fields in the tree level action \cite{Weinberg:2008hq, Burgess:2020tbq}. That is, one can substitute the tree level equations of motion  $R = - 8\pi G(\mu)T^{\rm bg}$ and $R_{\mu\nu} = 8\pi G(\mu)[T^{\rm bg}_{~~~\mu\nu} - \frac{1}{2}g_{\mu\nu}T^{\rm bg}]$ into the above, where
\eqn{}{T^{\rm bg\, \mu}_{~~~~~\nu}(X_B) = \delta^\mu_{~\nu} P^{\rm bg} - P^{\rm bg}_{~~~,X}\partial^\mu\psi_B\partial_\nu\psi_B}
is obtained from Eq. \ref{px} through variation with the background metric. The functional form $P^{\rm bg}(X_B) = X_B^2$ ensures that $T^{\rm bg} \equiv 0$, so that the regularized action takes the form:
\begin{equation}\label{full action 3}
	\left\langle S \right\rangle=  \int d^{4} x \sqrt{-g}\left[\frac{1}{2} M^2(\mu) R + \widetilde P^{\rm bg}(X_B,\mu)\right] + \left\langle S_{\rm gw}\right\rangle_{\rm  fin},
\end{equation}
where we've drop total derivatives and defined $M^{-2}(\mu) := 8\pi G(\mu)$, which we use interchangeably in what follows, and where 
\eqn{}{\widetilde P^{\rm bg}(X_B,\mu) := X_B^2 + 12X_B^4 \frac{\bar \alpha_2(\mu)}{M^4(\mu)}, }
with $\bar \alpha_2$ defined as in Eq. \ref{counterterms2}. We immediately notice that the stress tensor associated with the shifted matter sector $\widetilde P(X_B,\mu)$ is no longer traceless:
\eqn{trc}{\widetilde T^{\rm bg\, \mu}_{~~~~\,\mu} = -48X_B^4 \frac{\bar\alpha_2(\mu)}{M^4(\mu)}.}
This is because Einstein gravity is not conformally invariant, and therefore neither are the field equations governing gravitational waves even if the background is conformally flat \cite{Pandey:1983sp}. Hence the stress tensor for gravitational waves will not be exactly traceless unless $\bar\alpha_2 \equiv 0$, and this feature gets imported into the matter sector via operator redundancy at one loop, a point which we'll return to shortly. 

We now proceed to fix the finite parts of the relevant couplings via renormalization conditions. We recall the shifted tadpole condition, which is defined by the requirement that the background effective field $g_{\mu\nu}$ must be put on shell in all final expressions. That is, we demand that
\eqn{EOM1}{\frac{ 1}{8 \pi G(\mu)}\left(R_{\mu \nu} -\frac{1}{2} R g_{\mu \nu} \right) = \widetilde T^{\rm bg}_{\mu \nu} + \langle T^{\rm gw, fin}_{\mu \nu}\rangle.}
We first note that Newton's constant can only be fixed via a Cavendish type experiment, where we have knowledge of the masses whose strength of gravitational interactions we are attempting to fix. Let's say we do this at laboratory, i.e. mm scales, and impose the renormalization condition that $[8\pi G(\mu)]^{-1} \equiv M^2(\mu_*) \equiv M_{\rm pl}^2$ where the latter is given by the reduced Planck mass $M_{\rm pl}^2 = 2.435 \times 10^{18}$ GeV. From Eq. \ref{counterterms2} it therefore folows that 
\eqn{}{8\pi G_N(\mu) = \frac{1}{M_{\rm pl}^2}\left[1 + \frac{\alpha_1^{\rm F}(\mu) -  \alpha_1^{\rm F}(\mu_*) }{2\pi^2 M_{\rm pl}^2 } \right]^{-1},}
which can be used to express Eq. \ref{EOM1} in covariant form as
\eqn{gnfin}{G_{\mu\nu} = \frac{1}{M_{\rm pl}^2} \left[\widetilde T^{\rm bg}_{\mu \nu} + \langle T^{\rm gw, fin}_{\mu \nu}\rangle\right]\left[1 + \frac{\alpha_1^{\rm F}(\mu) -  \alpha_1^{\rm F}(\mu_*) }{2\pi^2 M_{\rm pl}^2 } \right]^{-1}.}
We see from the above how additive renormalization of Newton's constant is equivalent to the multiplicative renormalization of the matter sector via the tadpole condition. In the context of cosmology, however, this is a moot point as the factor in the square brackets above is precisely unity in the absence of any new light species. The reason for this is that at energy scales smaller than $\mu_*$ -- i.e. at scales bigger than laboratory scales, as with all cosmological observations -- the Newton's constant will no longer run\footnote{Even logarthmically, a fact that is immediately apparent with Pauli-Villars regularization, but obscured in dimensional and point-split regularization unless one performs threshold matching.} unless some new particle starts to propagate at a mass threshold lighter than the scale of the experiment. Given that the neutrino is the lightest massive particle that we know of, we can take the factor in the square brackets of Eq. \ref{gnfin} to be unity, although we leave it present in proceeding (see \cite{Negro:2024bbf} for further elaboration on this point).

In order to fix the remaining finite contribution appearing in $\bar \alpha_2$ in combination with those coming from $\langle T^{\rm gw, fin}_{\mu \nu}\rangle$, we have to appeal to additional observations. Before we do so, we note that $\langle T^{\rm gw, fin}_{\mu \nu}\rangle$ has a series of contributions that can be recursively obtained (cf. appendix \ref{Appendix B}). These contributions can be classed according to their scale factor dependence. We also note that the results of a mass independent regularization imply that the contribution Eq. \ref{trc} is evidently canceled by a compensating term from the state dependent part of $\langle T^{\rm gw, fin}_{\mu \nu}\rangle$ in the adiabatic vacuum \cite{Negro:2024bbf}. Regardless of this fact, in addition to admitting the possibility of other operators in the effective action from the presence of additional degrees of freedom, one simply extracts that part of the renormalized expression that scales as radiation and proceed accordingly.

Consider for example the measurement of the equation of state parameter $w$ during what we presume to be radiation domination. In principle, any other measurement of a dimensionless ratio will do (e.g. $H^2/M_{\rm pl}^2$) as what follows transcribes straightforwardly. Making such a measurement at the scale $\mu_{\rm R}$ results in the renormalization condition:
\eqn{eos}{3H^2_{\rm R}(3\omega_{\rm R} - 1) =   -48X_B^4 \frac{\bar\alpha_2(\mu_{\rm R})}{M^6_{\rm pl}} + \frac{\beta^{\rm F}(\mu_{\rm R})}{M^2_{\rm pl}},}
where $\beta^{\rm F}(\mu_{\rm R}) := \langle T^{\rm gw, fin}\rangle$. Regardless of how suppressed the right hand side may appear, it in principle fixes the remaining finite remainder in $\bar\alpha_2$ defined in Eq. \ref{counterterms2}, up to the state dependence of the terms that appear in $\beta^{\rm F}(\mu_{\rm R})$. If the right hand side cancels exactly, the effective background corresponds to radiation dominated expansion. If the right hand does not cancel, it would still correspond to radiation dominated expansion up to suppressed slow quenching terms\footnote{In practice, the best accuracy with which we can ever hope to constrain the left hand side of Eq. \ref{eos} means that the right hand side can only be taken in practice to be consistent with zero.}, which moreover, dilute much faster than radiation. Hence, the most one can conclude from the shifted tadpole condition Eq. \ref{gnfin} is:
\begin{eqnarray}
	\label{GWfin}
	3  H^2 &=& \frac{1}{M_{\rm pl}^2}\left(\widetilde \rho_{\rm bg} + \rho^{\rm rd}_{\rm gw, fin}\right)\left[1 + \frac{\alpha_1^{\rm F}(\mu) -  \alpha_1^{\rm F}(\mu_*) }{2\pi^2 M_{\rm pl}^2 } \right]^{-1} \approx \frac{\widetilde \rho_{\rm bg}}{M_{\rm pl}^2}\left(1 + \delta_{\rm Z}\right),  
\end{eqnarray}
where $\widetilde \rho_{\rm bg}$ corresponding to the time-time component of $\widetilde T^{\rm bg}_{\mu \nu}$, and with $\rho^{\rm rd}_{\rm gw, fin}$ denoting any (possibly vanishing) state dependent contributions from $\langle T^{\rm gw, fin}_{\mu \nu}\rangle$ that scales as radiation, and  $\delta_{\rm Z}$ is some constant by this definition. The latter defines the wavefunction renormalization of the otherwise unobservable bare thermal potential $\psi_B$:
\eqn{}{\psi := \left(1 + \delta_{\rm Z}\right)^{1/4}\psi_B,}
where now $\psi$ denotes a dressed quantity. Therefore one concludes that the net effect of the renormalization procedure is to simply mimic shifts in the definition of otherwise unaccessible quantities, a conclusion that can also be arrived at through dimensional regularization in the foliation specific formalism  \cite{Negro:2024bbf}.

\section{Summarizing remarks} \label{sec:6}

Whether one can meaningfully constrain vacuum tensor perturbations from cosmological observations such as $N_{\rm eff}$ bounds is intrinsically bound to the question of how one assigns a stress tensor to them, and how one regularizes the divergences that inevitably arise from the coincident limit of field bilinears that it samples. A typical expression that one can find in the literature (see e.g. \cite{Meerburg:2015zua, Maggiore2, Gasperini, Smith, Boyle, Lizarraga, Henrot, Cabass, Liu, Pagano, Li, Benetti, Berbig, Giare}) chooses to impose hard hard cutoffs on the energy density of gravitational waves obtained from the Isaacson form of the stress tensor. The result is expressions of the form 
\eqn{MH}{ \rho_{\mathrm{GW}} \simeq \frac{ A_{t}}{32 \pi  G_{ N}}\left(\frac{k_{\mathrm{UV}}}{k_*}\right)^{n_{ t}} \frac{1}{2 n_{ t}} \frac{1}{a^4} \propto \frac{1}{a^4}\left[\frac{1}{n_{ t}} + \log\frac{k_{\rm  UV}}{k_{*}} \right],}
where $k_*$ is some reference IR scale, and the approximation is only valid when $n_t \to 0$. It should be immediately apparent that no physical answer should depend on cutoffs, nor should one renormalize any quantity that presumes a prior scale separation in its definition. In a separate investigation \cite{Negro:2024bbf} we have shown how any attempts to extract $N_{\rm eff}$ bounds from vacuum tensor perturbations is inextricable from the process of background renormalization in a foliation specific formulation. Here, we have repeated this process in a covariant formalism, arriving at similar conclusions. We elaborate further on the consequences for cosmological observations, and what can meaningfully be interpreted from them in a follow up investigation \cite{NP}.


\appendix

\section{Details of Hadamard regularization}  \label{Appendix A}

In this appendix, we present further details as to how one can obtain the divergent contributions in Eq. \ref{eq:regularized action 1}. Regularizing the contributions of Eqs. \ref{eq:Actions 2.1} and \ref{eq:Actions 2.2} in order to obtain Eqs. \ref{regularized ghost action} and \ref{regularized graviton action}, boils down to regulating the following four terms:
\begin{eqnarray}\label{I}
	& \rm (I) : \quad & R_{\mu \alpha}  \lim _{\sigma^{\mu} \to 0} g_{\alpha^{\prime}}{}^{\alpha} \tilde{G}^{\mu \alpha^{\prime}}\\ \label{II}
	&\rm (II) : \quad & g_{\rho}{}^{ \gamma}\lim _{\sigma^{\mu} \to 0} g_{\tau^{\prime}}{}^{\tau} g_{\alpha^{\prime} \gamma} \nabla_{\tau} \nabla^{\tau^{\prime}} \tilde{G}^{\rho \alpha^{\prime}} \\ \label{III}
	& \rm (III) : \quad  & \mathcal{P}_{\mu \nu \alpha \beta} \lim _{\sigma^{\mu} \to 0} g_{\alpha^{\prime}}{}^{\alpha} g_{\beta^{\prime}}{}^{\beta} G^{\mu \nu \alpha^{\prime} \beta^{\prime}}  \\ \label{IV}
	& \rm (IV) : \quad  & \mathcal{Q}_{\mu \nu}{}^{ \gamma \delta} \lim _{\sigma^{\mu} \to 0} g_{\tau^{\prime}}{}^{\tau} g_{\alpha^{\prime} \gamma} g_{\beta^{\prime} \delta} \nabla_{\tau} \nabla^{\tau^{\prime}} G^{\mu \nu \alpha^{\prime}\beta^{\prime}},
\end{eqnarray}
where we've defined
\begin{eqnarray}\label{Q e P}
	&&\mathcal{P}_{\mu \nu \gamma \delta} \equiv R_{\gamma \mu \delta \nu}+g_{\delta \nu} R_{\mu \gamma}-g_{\gamma \delta} R_{\mu \nu}-\frac{1}{2} R \: g_{\mu \gamma} g_{\nu \delta} +\frac{1}{4} R \:g_{\mu \nu} g_{\gamma \delta},\\
	\nonumber &&\mathcal{Q}_{\mu \nu \gamma \delta} \equiv -\frac{1}{2} g_{\mu \gamma} g_{\nu \delta}+\frac{1}{4} g_{\mu \nu} g_{\gamma \delta}.
\end{eqnarray}
By expanding the Hadamard Green's function of Eq. \ref{eq:Hadamard propagators} using the Taylor expansions in Eq. \ref{eq:Taylor expansions}, we obtain
\begin{eqnarray}\label{eq:Hadamard function expanded}
	G^{\rho \sigma \alpha^{\prime} \beta^{\prime}}\left(x, x^{\prime}\right)&=&\frac{1}{4 \pi^{2}}\left[\frac{\Delta^{1 / 2}}{\sigma}\left(g^{\alpha^{\prime} (\rho}g^{\sigma) \beta^{\prime}} \right)+V^{\rho \sigma \alpha^{\prime} \beta^{\prime}} \ln (\mu^2\sigma)+W^{\rho \sigma \alpha^{\prime} \beta^{\prime}}\right]\\
	\nonumber	&=&\frac{1}{4 \pi^{2}}\left[\frac{\Delta^{1 / 2}}{2 \sigma}\left(g^{\alpha^{\prime} \rho} g^{\sigma \beta^{\prime}}+g^{\alpha^{\prime} \sigma} g^{\rho \beta^{\prime}}\right)+g^{\alpha^{\prime}}{}_{\alpha} g^{\beta^{\prime}}{}_{\beta} v_{0}^{\rho \sigma\alpha\beta} \ln (\mu^2\sigma)+g^{\alpha^{\prime}}{}_{\alpha} g^{\beta^{\prime}}{}_{\beta} v_{0}^{\rho \sigma \alpha \beta}{}_{\gamma}\sigma^{\gamma} \ln (\mu^2\sigma)\right.\\
	\nonumber	&&\left.+\frac{1}{2} g^{\alpha^{\prime}}{}_{\alpha} g^{\beta^{\prime}}{}_{\beta} v_{0}^{\rho \sigma \alpha \beta}{}_{\gamma \varepsilon}  \sigma^{\gamma} \sigma^{\varepsilon} \ln (\mu^2\sigma)+\frac{1}{2} g^{\alpha^{\prime}}{}_{\alpha} g^{\beta^{\prime}}{}_{\beta} v_{1}^{\rho \sigma \alpha \beta} \sigma_{\gamma} \sigma^{\gamma} \ln (\mu^2\sigma)+g^{\alpha^{\prime}}{}_{\alpha} g^{\beta^{\prime}}{}_{\beta} w_{0}^{\rho \sigma\alpha\beta} \right.\\
	\nonumber	&&\left. +g^{\alpha^{\prime}}{}_{\alpha} g^{\beta^{\prime}}{}_{\beta} w_{0 }^{\rho \sigma\alpha\beta}{}_{\gamma} \sigma^{\gamma}+\frac{1}{2} g^{\alpha^{\prime}}{}_{\alpha} g^{\beta^{\prime}}{}_{\beta} w_{0 }^{\rho \sigma\alpha\beta}{}_{\gamma \tau} \sigma^{\gamma} \sigma^{\tau} +\frac{1}{2} g^{\alpha^{\prime}}{}_{\alpha} g^{\beta^{\prime}}{}_{\beta} w_{1}^{\rho \sigma\alpha\beta} \sigma^{\gamma} \sigma_{\gamma} \right],\\
	\tilde{G}^{\mu \alpha^{\prime}}\left(x, x^{\prime}\right)&=&\frac{1}{4 \pi^{2}}\left[\frac{\Delta^{1 / 2}}{\sigma} g^{\mu \alpha^{\prime}}+\tilde{V}^{\mu \alpha^{\prime}} \ln (\mu^2\sigma)+\tilde{W}^{\mu  \alpha^{\prime} }\right]\\
	\nonumber	&=&\frac{1}{4 \pi^{2}}\left[\frac{\Delta^{1 / 2}}{\sigma} g^{\mu \alpha^{\prime}}+g^{\alpha^{\prime}}{}_{\alpha} \tilde{v}_{0}^{\mu \alpha} \ln (\mu^2\sigma)+g^{\alpha^{\prime}}{}_{\alpha} \tilde{v}_{0}^{\mu \alpha}{}_{\gamma}  \sigma^{\gamma} \ln (\mu^2\sigma)+\frac{1}{2} g^{\alpha^{\prime}}{}_{\alpha} \tilde{v}_{0 }^{\mu \alpha}{}_{\gamma \varepsilon} \sigma^{\gamma} \sigma^{\varepsilon} \ln (\mu^2\sigma)\right.\\
	\nonumber	&&\left.+\frac{1}{2} g^{\alpha^{\prime}}{}_{\alpha}  \sigma_{\gamma} \sigma^{\gamma} \tilde{v}_{1}^{\mu \alpha} \ln (\mu^2\sigma) +g^{\alpha^{\prime}}{}_{\alpha}  \tilde{w}_{0}^{\mu \alpha}+g^{\alpha^{\prime}}{}_{\alpha} \tilde{w}_{0 }^{\mu \alpha}{}_{\gamma} \sigma^{\gamma}+\frac{1}{2} g^{\alpha^{\prime}}{}_{\alpha}  \tilde{w}_{0 }^{\mu \alpha}{}_{\gamma \tau} \sigma^{\gamma} \sigma^{\tau} +\frac{1}{2} g^{\alpha^{\prime}}{}_{\alpha} \tilde{w}_{1}^{\mu \alpha} \sigma^{\gamma} \sigma_{\gamma} \right],
\end{eqnarray}
where higher orders in powers of $\sigma$ vanish in the limit $\sigma^\mu \to 0$, and the tensors contributing to the divergent part $\{v_{0}^{\rho \sigma\alpha\beta}$, $v_{0}^{\rho \sigma \alpha \beta}{}_{\gamma}$, $v_{0}^{\rho \sigma \alpha \beta}{}_{\gamma \varepsilon}$, $v_{1}^{\rho \sigma \alpha \beta} $, $\tilde{v}_{0}^{\rho \alpha}$, $\tilde{v}_{0}^{\rho \alpha}{}_{\gamma }$, $\tilde{v}_{0 }^{\rho \alpha}{}_{\gamma \varepsilon}$ and $\tilde{v}_{1}^{\rho \alpha}\}$ are found by iteratively solving the equations of motion for the propagators and are given in section A.3 of \cite{GW}.

Eqs. \ref{I} and \ref{III} are then regularized by subtracting the divergent terms of the expansions in Eq. \ref{eq:Hadamard function expanded} with the appropriate counterterms. These divergences are given by
\begin{eqnarray}\label{I and III}
	\rm (I)_{\rm div} : \: &&-\frac{1}{16 \pi^2} \lim _{\sigma^{\mu} \to 0} \left[\frac{1}{\sigma} R+\ln (\mu^2\sigma) \left( -\frac{1}{12} R^2 -\frac{1}{2}R_{\mu \nu} R^{\mu \nu}\right) \right] \\ \nonumber
	\rm (III)_{\rm div} : \: && -\frac{1}{16 \pi^2} \lim _{\sigma^{\mu} \rightarrow 0} \left[-3 \frac{1}{\sigma} R + \ln (\mu^2\sigma) \left( \frac{3}{2} R_{\mu \nu} R^{\mu \nu}-R^{2}-\frac{1}{2} R_{\mu \nu \rho  \sigma} R^{\mu \nu \rho \sigma}-\frac{1}{2} R_{\mu \rho \nu  \sigma} R^{\mu \nu \rho \sigma} \right)  \right],
\end{eqnarray}
and the requisite counterterms are readily identified. Regularizing Eq. \ref{II} and Eq. \ref{IV} is less straightforward, as in order to regularize $\nabla_{\tau} \nabla^{\tau^{\prime}} \tilde{G}^{\rho \alpha^{\prime}}$ and $\nabla_{\tau} \nabla^{\tau^{\prime}} G^{\mu \nu \alpha^{\prime}\beta^{\prime}}$ we need to sequentially:
\begin{enumerate}
	\item Compute the derivative of the Hadamard Green's functions using the expansions in \ref{eq:Hadamard function expanded} and keeping the terms that are divergent in the limit $\sigma^{\mu} \to 0$.
	\item Expand the result in powers of $\sigma^{\mu}$ using the endpoint expansions in Ref. \cite{Adler:1976jx}.
	\item Use the averages in Eq. \ref{eq:average} to obtain a direction independent result.
\end{enumerate}
Following these steps, one obtains
\begin{equation}\label{II and IV}
	\begin{aligned}
		\rm (II)_{\rm div} : \: & -\frac{1}{16 \pi^2} \lim _{\sigma^{\mu} \to 0} \left[\frac{2}{\sigma} R  + \ln(\mu^2\sigma)) \left(- \frac{1}{2} R_{\mu \nu} R^{\mu \nu}-\frac{1}{12}R^{2}+\frac{1}{12} R_{\mu \nu \rho  \sigma} R^{\mu \nu \rho \sigma}-\frac{1}{12} R_{\mu \rho \nu  \sigma} R^{\mu \nu \rho \sigma} - \frac{5}{12} \Box R \right) \right] \\
		\rm (IV)_{\rm div} :  \: &-\frac{1}{16 \pi^2} \lim _{\sigma^{\mu} \to 0} \left[\frac{11}{6} \frac{1}{\sigma} R +  \ln(\mu^2\sigma) \left( - \frac{1}{2} R_{\mu \nu} R^{\mu \nu}+\frac{3}{4}R^{2}+\frac{1}{2} R_{\mu \nu \rho  \sigma} R^{\mu \nu \rho \sigma}+\frac{1}{2} R_{\mu \rho \nu  \sigma} R^{\mu \nu \rho \sigma} - \frac{13}{24} \Box R  \right) \right].
	\end{aligned}
\end{equation}
In summary, the divergent contributions in Eq. \ref{divc} are given by\footnote{The extra minus in front of the ghost terms accounts for the different statistics.}
\begin{eqnarray}\label{regularized action app}
	\left\langle S\right\rangle_{\rm div}  &=& \int d^{4} x \sqrt { - g } \left[\Big( \rm (III)_{\rm div} + \rm (IV)_{\rm div} \Big)- \Big( -\rm (I)_{\rm div} - \rm (II)_{\rm div} \Big) \right]\\
	\nonumber	&=&\frac{1}{4 \pi^2} \lim _{\sigma^{\mu} \to 0}  \int d^{4} x \sqrt { - g } \left[\frac{11}{6}\frac{1}{\sigma}  R+ \ln (\mu^2\sigma) \left(\frac{1}{6} R_{\mu \nu} R^{\mu \nu} -\frac{11}{24} R^2 -\frac{23}{24} \Box R\right) \right] 
\end{eqnarray}
where we've used the relevant Bianchi identity to obtain $2 R_{\mu \nu \rho  \sigma} R^{\mu \rho \nu \sigma}=R_{\mu \nu \rho  \sigma} R^{\mu \nu \rho \sigma}$, and the Gauss-Bonnet theorem to rewrite the Riemann squared terms in terms of the Ricci tensor and scalar.

\section{Finite contributions} \label{Appendix B}

In this appendix, we detail the recursion relations that determine the finite remainder given by $\left\langle S_{\rm gw}\right\rangle_{\rm fin} := \left\langle S_{\rm gw}\right\rangle - \left\langle S_{\rm gw}\right\rangle_{\rm div} $ as it appears in Eq. \ref{full action 2} for completeness. We first note that we can separate the finite contribution of the Hadamard regularized action into parts that are uniquely determined by the background geometry -- i.e. the terms determined by the bitensors $V^{\mu \nu \alpha^{\prime} \beta^{\prime}}$ and $\tilde{V}^{\mu \alpha^{\prime}} $ which we denote as $\left\langle S_{\rm gw}\right\rangle_{\rm fin}^{\rm bg}$ -- and those that depend on the state, determined by the bitensors $W^{\mu \nu \alpha^{\prime} \beta^{\prime}}$ and $\tilde{W}^{\mu \alpha^{\prime}}$, which we denote as $\left\langle S_{\rm gw}\right\rangle_{\rm fin}^{\rm sd}$.

In order to compute $\left\langle S_{\rm gw}\right\rangle_{\rm fin}^{\rm sd}$, we follow the procedure of appendix \ref{Appendix A} by considering the state dependent terms of the Taylor expansions of Eq. \ref{eq:Hadamard function expanded} -- $w_{0}^{\rho \sigma\alpha\beta}$, $w_{0}^{\rho \sigma \alpha \beta}{}_{\gamma}$, $w_{0}^{\rho \sigma \alpha \beta}{}_{\gamma \varepsilon}$, $w_{1}^{\rho \sigma \alpha \beta} $, $\tilde{w}_{0}^{\rho \alpha}$, $\tilde{w}_{0}^{\rho \alpha}{}_{\gamma }$, $\tilde{w}_{0 }^{\rho \alpha}{}_{\gamma \varepsilon}$ and $\tilde{w}_{1}^{\rho \alpha}$. We find that the finite contribution to the terms (I), (II), (III) and (IV) defined in Eqs. \ref{I} - \ref{IV} are given by:
\begin{eqnarray}\label{I and III}
\rm (I)^{\rm sd}_{\rm fin} : \: && \frac{i}{8 \pi^2} R_{\mu \alpha} \tilde{w}_{0 }^{\mu \alpha} \\ \nonumber
\rm (II)^{\rm sd}_{\rm fin} : \: && \frac{i}{8 \pi^2} \left[ -\nabla_\tau \tilde{w}_{0}^{\mu  }{}_{\mu}{}^{\tau} -\tilde{w}_{0 }^{\mu }{}_{\mu }{}^{\tau}{}_{\tau} -4 \tilde{w}_{1}^\mu{}_{\mu } \right] \\ \nonumber
\rm (II)^{\rm sd}_{\rm fin} : \: &&\frac{i}{8 \pi^2} \mathcal{P}_{\mu \nu  \alpha \beta} w_0^{\mu \nu  \alpha \beta} \\ \nonumber
\rm (IV)^{\rm sd}_{\rm fin} :  \: && \frac{i}{8 \pi^2} \left[ \mathcal{Q}_{\mu \nu}{}^{  \gamma \delta} \left( -\nabla_{\tau} w_{0 }^{\mu \nu }{}_{\gamma \delta}{}^{\tau} -w_{0}^{\mu \nu }{}_{ \gamma \delta }{}^{\tau}{}_{\tau}-4 w_{1 }^{\mu \nu}{}_{\gamma \delta}\right) \right]
\end{eqnarray}
so that $\left\langle S_{\rm gw}\right\rangle_{\rm fin}^{\rm sd}$ can be expressed as
\begin{eqnarray}\label{finite action app}
		\left\langle S\right\rangle_{\rm fin}^{\rm sd}  &=& \int d^{4} x \sqrt { - g } \left[\Big( \rm (III)^{\rm sd}_{\rm fin} + \rm (IV)^{\rm sd}_{\rm fin} \Big)- \Big( -\rm (I)^{\rm sd}_{\rm fin} - \rm (II)^{\rm sd}_{\rm fin}\Big) \right]\\ \nonumber
		&=&\frac{1}{4 \pi^2}  \int d^{4} x \sqrt { - g } \left[ \mathcal{Q}_{\mu \nu}{}^{  \gamma \delta} \left( -\nabla_{\tau} w_{0 }^{\mu \nu }{}_{\gamma \delta}{}^{\tau} -w_{0}^{\mu \nu }{}_{ \gamma \delta }{}^{\tau}{}_{\tau}-4 w_{1 }^{\mu \nu}{}_{\gamma \delta}\right) +\mathcal{P}_{\mu \nu  \alpha \beta} w_0^{\mu \nu  \alpha \beta}+R_{\mu \alpha} \tilde{w}_{0 }^{\mu \alpha} \right.\\ \nonumber
		&& \left. -\nabla_\tau \tilde{w}_{0}^{\mu  }{}_{\mu}{}^{\tau} -\tilde{w}_{0 }^{\mu }{}_{\mu }{}^{\tau}{}_{\tau} -4 \tilde{w}_{1}^\mu{}_{\mu }  \right].
\end{eqnarray}
We note that we can obtain the Taylor coefficients of $W^{\mu \nu \alpha^{\prime} \beta^{\prime}}_{1} $ in terms of the Taylor coefficients of $W^{\mu \nu \alpha^{\prime} \beta^{\prime}}_{0} $ (we focus on the graviton contribution, but a similar procedure will give us the analogous Taylor coefficients for the ghost contributions). By iteratively solving order by order in $\sigma^\mu$ in the equation of motion for the propagator, we find (cf. \cite{GW} for more details):
\begin{eqnarray}\label{recursion relation Wn}
		&& n(n+1) W_{n}^{\mu \nu}{}_{\alpha^{\prime} \beta^{\prime}}+n W_{n}^{\mu \nu}{}_{ \alpha^{\prime} \beta^{\prime} ; \rho} \sigma^{\rho}-n W_{n }^{\mu \nu}{}_{\alpha^{\prime} \beta^{\prime}} \Delta^{-1 / 2} \Delta_{; \rho}^{1 / 2} \sigma^{\rho}+(2 n+1) V_{n }^{\mu \nu}{}_{\alpha^{\prime} \beta^{\prime}}+V_{n }^{\mu \nu}{}_{\alpha^{\prime} \beta^{\prime} ; \rho} \sigma^{\rho} \\  \nonumber
		&&-V_{n }^{\mu \nu}{}_{\alpha^{\prime} \beta^{\prime}} \Delta^{-1 / 2} \Delta_{; \rho}^{1 / 2} \sigma^{\rho}+\frac{1}{2} D_{\rho \sigma}{}^{\mu \nu} \:  W_{n-1 }^{\rho \sigma}{}_{\alpha^{\prime} \beta^{\prime}}=0,
\end{eqnarray}
where
\begin{eqnarray}
		&&D_{\mu \nu}{}^{\alpha \beta}=\Box g_\mu^{(\alpha} \: g_\nu^{\beta)} -P_{\mu \nu}{}^{\alpha \beta} , \\ \nonumber
		&& P_{\mu \nu}{}^{\alpha \beta}=-2 R_{(\mu}{ }^{\alpha}{ }_{\nu)}{ }^{\beta}+\frac{1}{2}g_{\mu \nu} R^{\alpha \beta}+\frac{1}{2}g^{\alpha \beta} R_{\mu \nu}-\frac{1}{4} R g_{\mu \nu} g^{\alpha \beta} +\frac{1}{2} R g_{(\mu}{ }^{\alpha} g_{\nu)}{ }^{\beta}.
\end{eqnarray}		
By specifying the recursion relation Eq. \ref{recursion relation Wn} for $n=1$ and expanding at the $0^{th}$ order in $\sigma$ we obtain $w_{1}^{\mu \nu}{ }_{\alpha \beta}$ as a function of the Taylor coefficients of $W^{\mu \nu \alpha^{\prime} \beta^{\prime}}_{0} $
\begin{equation}\label{recursion relation w1}
	2 w_{1}^{\mu \nu}{ }_{\alpha \beta} = -3 v_{1}^{\mu \nu}{ }_{\alpha \beta} -\frac{1}{2} g_\rho^{(\mu} \: g_\sigma^{\nu)} \left[ \Box w_{0}^{\rho \sigma}{ }_{\alpha \beta}+\nabla_{\tau} w_{0}^{\rho \sigma}{ }_{\alpha \beta}{ }^\tau +\frac{1}{2} w_{0}^{\rho \sigma}{ }_{\alpha \beta}{ }^\tau{ }_\tau  \right] + \frac{1}{2}   P^{\mu \nu}{ }_{\rho \sigma} \: w_{0}^{\rho \sigma}{ }_{\alpha \beta} . 
\end{equation}
In the above $w_{0}^{\rho \sigma\alpha\beta}$, $w_{0}^{\rho \sigma \alpha \beta}{}_{\gamma}$ and $w_{0}^{\rho \sigma \alpha \beta}{}_{\gamma \varepsilon}$ are the `initial' inputs for the recursion relations corresponding to the specifics of the state. The adiabatic vacuum by definition is invariant under the symmetries of the background geometry, and so all the initial state dependent inputs must themselves be constructed out of geometric invariants. By scanning through possibilities by rank, one concludes that the latter will also result in the generation of a handful of terms that redshift as radiation (if not vanish outright \cite{Negro:2024bbf}\footnote{One can proceed similarly for the finite background dependent contributions, and the result will be a similar set of recursion relations initial inputs that will be degenerate with the finite remainders to the counterterms.}) along with a series of additional slow quenching terms that decay much faster. 

We close by addressing how one can compare the results of this investigation with the gauge fixed, foliation specific treatment of \cite{Negro:2024bbf}. We note that in fixing de Donder gauge with the Faddeev Popov method, we began with the action for a rank-2 symmetric tensor field and gauge fixed via the gauge breaking term ($\nabla_\mu h^{\mu}_{\nu} = \frac{1}{2}\nabla_\nu h$, thus eliminating four degrees of freedom with residual gauge symmetry left over) with the vector ghosts subtracting the remaining 4 spurious degrees of freedom. The ghost and gauge fixing terms in Eq. \ref{eq:sgw} possess the same properties as the eight spurious degrees of freedom present in the fully diffeomorphism invariant action for a rank-2 symmetric tensor field, but with fermionic statistics that subtracts them from all on shell quantities. In order to arrive at a fully gauge fixed action in terms of only the transverse traceless polarizations of the graviton, one would have to determine the ghost propagator in terms of the graviton propagator using the generalization of the Ward identities discussed in \cite{GW}, but now evaluated on a background that does not correspond to a vacuum spacetime. Although this is beyond the scope of the present investigation, to do so to completion given the residual symmetries of FRLW spacetimes would be a practical and important computation to follow up on.

\end{document}